# Artificial Immune Systems Metaphor for Agent Based Modeling of Crisis Response Operations

Khaled M. Khalil[1], M. Abdel-Aziz[1], Taymour T. Nazmy[1],
Abdel-Badeeh M. Salem[1]

[1]Faculty of Computer and Information Science Ain shams University Cairo, Egypt
kmkmohamed@gmail.com, mhaziz67@gmail.com, ntaymoor@yahoo.com,
absalem@asunet.shams.edu.eg

**Abstract.** Crisis response requires information intensive efforts utilized for reducing uncertainty, calculating and comparing costs and benefits, and managing resources in a fashion beyond those regularly available to handle routine problems. This paper presents an Artificial Immune Systems (AIS) metaphor for agent based modeling of crisis response operations. The presented model proposes integration of hybrid set of aspects (multi-agent systems, built-in defensive model of AIS, situation management, and intensity-based learning) for crisis response operations. In addition, the proposed response model is applied on the spread of pandemic influenza in Egypt as a case study.

## 1 Introduction

The challenge of crisis response is reducing the influence crises cause to society, the economy, and the lives of individuals and communities. This challenge is extreme in several dimensions. The demand is highly diverse and largely unpredictable in terms of location, time, and specific resources needed. Moreover, the urgency associated with crisis has many implications, such as the need to rapidly identify information about the developing situation, and to have the capability to make good decisions in the face of an inevitable degree of uncertainty and incompleteness of information. An efficient crisis response is of paramount importance, because if not responded to promptly and managed properly, even a small mishap could lead to a very big catastrophe with significantly severe consequences. Being equipped with a profusion of resources does not ensure a successful response to the crisis situation. Thus, the key to the successful response necessitates an effective and expedited allocation of the requested resources to the emergency locations. Such complexity suggests the use of intelligent agents for adaptive real-time modeling of the crisis response operations [16]. Multi-agent Systems are computational systems where software agents cooperate or compete with each other to achieve an individual or collective task [30].

In order to build multi-agent architecture, and increase the effectiveness of the crisis response operations, a similar metaphor is required that mimics the crisis response operations. AIS metaphor is selected in this study. AIS are a computational systems inspired by the principles and processes of the biological immune system [5] [9]. AIS represent an area of vast research over the last few years. For example, developed AIS in a variety of domains, such as machine learning [14], anomaly detection [4] [10], data mining [21], computer security [20] [6], adaptive control [24] and fault detection [7]. The biological immune system is a robust, complex, adaptive system that defends the body from foreign pathogens. It is able to categorize all cells (or molecules) within the body as self-cells or non-self cells. It does this with the help of a distributed task force that has the intelligence to take action from a local and also a global perspective using its network of chemical messengers for communication [6]. A more detailed overview of the immune system can be found in many textbooks [23] [26]. The immune system combines a priori knowledge with the adapting capabilities of a biological immune system to provide a powerful alternative to currently available techniques for pattern recognition, learning and optimization [22]. It uses several computational models and algorithms such as Bone Marrow Model, Negative Selection Algorithm, and Clonal Selection Algorithm [13].

In this paper we propose a multi-agent based model for crisis response. The proposed model architecture and operations process are adopted from AIS. Then the proposed response model is applied on controlling pandemic influenza in Egypt. Section 2 provides the proposed response model, while section 3 presents design of the proposed model for pandemic influenza in Egypt. Section 4 includes experiments. Finally, section 5 includes conclusions.

## 2  The Proposed Response Model

The view of the biological immune system provides the basis for a representation of AIS as systems of autonomous agents which exist within a distributed and compartmentalized environment [29]. In what follows, we present the multi-agent model based on the AIS metaphor for crisis response operations.

### 2.1  Proposed Hierarchical Architecture for Multi-Agent Response Model

The architecture of the AIS can be abstracted into hierarchy of three levels (cells, tissue, and host) (see Fig. 1). Cells are able to interact with their environment and communicate and coordinate their behavior with other cells by synthesizing and responding to a range of molecules. Cells within the body aggregate to form tissue, such as muscle or connective tissue. Tissues themselves combine to form hosts, such as the heart, brain, or thymus. Hosts work together to form the immune system.

The proposed multi-agent architecture follows the same hierarchical architecture of the biological immune systems with mapping of the functionalities of cells to agents and adopting crisis response domain attributes and operations levels (operational, tactical and strategic levels [2] [3]) (see Table 1). Pathogens represent source of danger to the body entity, in which immune systems antibody cells tries to detect and kill. Pathogens are mapped to danger sources or undesired situations in the crisis domain, in which agents have to detect and overcome. Cells are represented by agents working as first responders and voluntaries. Cells contain different type of receptors which affect their capability to match pathogens and to kill them. Thus, receptors are mapped to agents skills or resources required to overcome danger. Agents are working in groups belonging to certain organization (tissue) which provide help by other agents teams dedicated in the tactical level. Host represents the grouping of different tissues working together. This can be mapped to emergency operations center (EOC) of different working divisions and each division contains specialized teams.

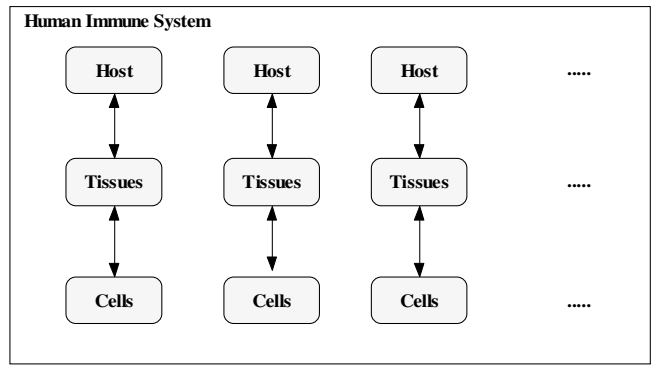

**Fig. 1.** Hierarchical architecture of human immune system

**Table 1.** Mapping biological immune system levels to crisis response operations and levels

| Biological Immune System Level | Crisis Response Level | Crisis Domain |
| --- | --- | --- |
| Pathogens | - | Danger sources – harmful situations |
| Cellular | Operational | First Responders and Voluntaries |
| Tissue | Tactical | Helper Agents staff |
| Host | Strategic | Emergency Operations Centers |
| System | - | Human Society and Important Properties |

### 2.2 AIS Operational Architecture for Multi-Agent Model

However, the AIS layered structure which is adopted in previous section is not complete from the conceptual framework perspective, which is required to allow effective algorithms to be developed [28]. Brownlee [1] said that "The acquired immune system provides a general pattern recognition and defense system that modifies itself, adapting to improve its capability with experience".

**Decision Making Process of AIS (Conceptual Model Formalization)**
The effectiveness of the system is due to a set of internal strategies to cope with pathogenic challenges. Such strategies remodel over time as the organism develops, matures, and then ages. Towards determining the decision making process of AIS, rational reconstructions approach is used. Rational reconstructions operate so as to transform a given problematic philosophical scientific account-particularly of a terminological, methodological or theoretical entity-into a similar, but more precise, consistent interpretation [8]. Proposed rational reconstruction of AIS follows the same steps of Grant et al. work [12]. Grant et al. steps include: definition of requirements of the model, definition of the top-level use-cases, selection of notation, formalization of the process model by walking through use-cases, implementation, and evaluation.

Definition of requirements of the proposed response model follows crisis response systems design requirements [17] [18]. The following requirements were defined as:
- The model processes are concurrent.
- Support multiple instances of agents.
- Agents located in different layers should share required information only.
- Allow continuous monitoring of the situation and available resources.
- Permit relationships between agents to be collaborative.
- Allow separate behavior of each agent based on its role and objectives.
- Allow removal and adding of new agents and components at run time.
- Allow continuous planning/re-planning.
- Integrate planning and learning processes.
- Each process defined can be done by one or more agents. Agents with different roles differ by their allocated processes.

Definition of the top-level use cases is as follows (see Fig. 2):
- Use-case (0): no change in environment. This use-case applies when no pathogens found in the environment.
- Use-case (1): antibody found a pathogen. This use-case applies when the antibody finds a pathogen in the environment.
- Use-case (2): antibody receptors detect the required response for pathogens. This use-case applies when the antibody receptors match the pathogen, and can provide required response.
- Use-case (3): antibody receptors cannot detect the type of the pathogen thus failed to provide required response. This use-case applies when the antibody failed to match the pathogen and failed to provide response.
- Use-case (4): antibody cell asks for help for handling the unknown pathogen. This use-case applies when antibody failed to response to the pathogen. Antibody sends signals to activate the adaptive response.
- Use-case (5): adaptive cells mutate to match the pathogen and generate required receptors. This use-case applies when the adaptive cells mutate to match the pathogen and generate the required receptors.
- Use-case (6): required receptors are cloned to be applied to pathogen. This use-case applies when the required receptors for response are cloned to provide response to pathogen.
- Use-case (7): successful response is sustained as memory cells. This use-case applies when a successful response is executed; the receptors are stored in memory cells for later usage. Go to use-case (0).
- Use-case (8): failed response is ignored. This use-case applies when a failed response is gained; the receptors are ignored and not stored in memory. Go to use-case (5).

Selection of Notation includes selection of process notation for representing the process model. Integrated DEFinition Methods Technique (IDEFS0) [15] represents a common notation used in process modeling which is highly suited to specifying systems in terms of functional processes [12].

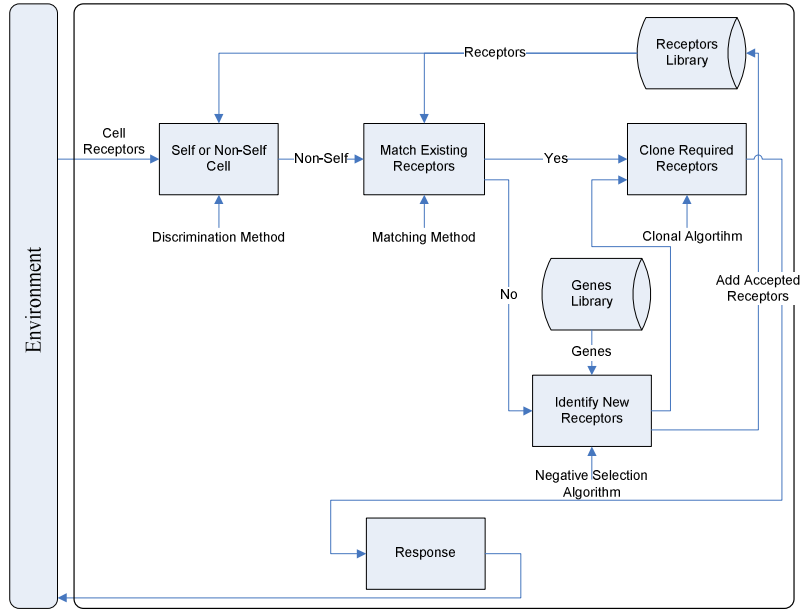

**Fig. 2.** Rationally reconstructed AIS model, formalized using SADT notation

The rationally reconstructed model (see Fig. 2) can be explained by starting at the environment and considering the activity of a typical Cell. Antibody cell checks other cells in the environment. Antibody cell determines if the examined cell is self or non-self based on the available receptors in the system using specific discrimination method. If cell is identified as non-self, cell immediately tries to match the receptors of the cell with existing receptors for response using matching method. If response receptors are found, clonal of the antibodies using clonal algorithm is applied to attack the pathogen cell. In case of new pathogen receptors not currently recognized, cell activates the adaptive response mechanism. Cells in the adaptive response mechanism mutate to match the non-self receptors using available genes library and negative selection algorithm. When reach an acceptable receptors form, the generated antibodies are cloned to provide response to the pathogen cell. Then, the newly generated receptors are added to the receptors library.

Table 2 shows comparison between the AIS rational reconstructed model and other process models such as OODA, RPDM, and Rasmussen models. Comparison criteria and OODA, RPDM, and Rasmussen models values are presented by Grant et al. [11].

**Table 2.** Comparing proposed AIS, OODA, RPDM, and Rasmussen models

| Criteria/Process Model | OODA | Rasmussen | RPDM | AIS Model |
|---|---|---|---|---|
| Control Loop | √ | √ | √ | √ |
| Detailed | × | × | √ | √ |
| Tempo (fast decision making) | √ | × | × | × |
| Planning | × | √ | × | √ |
| Learning | × | × | × | √ |

Formalizing the process model by walking through use-case. Use case (8) represents failing to response to pathogen cells. To formalize the process model, steps of the use case are presented as follows (see Fig. 3):

1. Antibody cell examines other cells looking for pathogens.
2. Antibody cell detects that cell is non-self.
3. Antibody cell tries to find proper receptors to kill the pathogen.
4. Antibody cell failed to response.
5. Antibody cell activates the adaptive system to generate proper receptors.
6. Adaptive cells mutate to match the pathogen and generate required receptors.
7. Required receptors are cloned and response is provided.
8. Failed response is reported as the pathogen is detected again which backs to step 1.

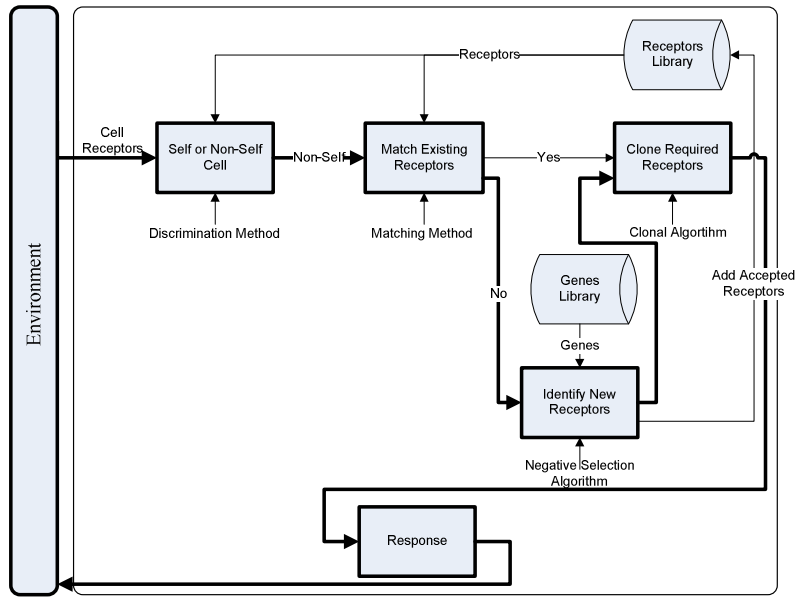

**Fig. 3.** Walk through use case (8), failed response to pathogen

**Mapping AIS Operational Model to Crisis Response Multi-Agent Model**

After definition of the conceptual model of the AIS, mapping the proposed model to multi-agent model for crisis response is a straight forward process (see Tables 3 and 4). An agent can examine environment searching for danger sources or undesired feature. This can be mapped to operational agents. When agent finds an un-desired situation in environment, agent tries to handle the problem using available procedures. If the situation exceeds the available routine procedures, agents ask for help from tactical agents. Tactical agents check available memory for similar situations and check if the old experienced situation can be adopted for the current situation or not. If an old situation matches the current state of the environment, apply the course of actions coupled with the experienced situation. Otherwise, tactical agent asks for decision making (strategic) agent help. Using nearest matched situation, decision making agent mutates different course of actions to handle the current situation till reach an acceptable course of actions (intensity-based learning [25]). Decision making agent allocates course of actions to tactical agents to be deployed (cloned). During the execution of actions, operational agents report status of tasks execution, and in case of failed task re-planning is presented. Finally, generation and death of agents are related to the application domain of the proposed model. In crisis response death of cells and generation of new cells can be mapped to deployment of effective actions and neglecting others, or removal and adding new responders to the response field.

**Table 3.** Mapping AIS operational model to crisis response multi-agent model

| Biological Immune System Level | Crisis Response Domain |
|---|---|
| Cells | Operational agent |
| Helper Cells | Tactical agents |
| Cells mutation | Decision making agents |
| Memory Cells | Case Memory |

**Table 4.** Mapping AIS Operational Model Processes to agents' roles

| Biological Immune System Process | Crisis Domain | Agent Role |
|---|---|---|
| Self or non-self discrimination | Reporting un-desired situation | Operational |
| Matching existing receptors | Situation Recognition | Tactical |
| Identify new receptors | Planning (Situation Assessment) | Decision making |
| Clone required receptors | Allocation of plan tasks | Tactical |
| Response | Executing plan tasks | Operational |
| Cells generation and death | Deployment of effective actions and neglecting non-effective actions | |

## 3   Crisis Response to Pandemic Influenza in Egypt

This section presents the design, and implementation of the proposed response model for pandemic influenza in Egypt. Agent environment includes two parts: the pandemic model which is implemented in the previous work [19], and the available resources (control strategies or actions). Agent retrieves the current pandemic situation based on the agents' health states. Each control strategy is represented by (resource type, amount, cost, from/to date, and efficiency [19]). Agents' roles argue that each agent has its own profile and roles which specify its responsibilities and skills. For example decision making agent has the role of making decisions and adding new memory cases, while tactical agent has the role of processing information and allocating tasks to operational agents. Table 5 shows different agent roles, role responsibilities, and number of agents allowed per emergency operations center. Added here a new role titled tactical communication agent. Actually tactical communication agent is a tactical agent which is specialized for managing communication among the EOC parties. Tactical communication agent has to receive reports from operation agents, send reports to other tactical agents, deliver reports to crisis decision maker, and deliver plans back to the tactical agents.

Crisis decision making agent follows the BDI (Belief-Desire-Intention) cognitive model [27]. While, tactical and operational agents are recognized as reactive helper agents for the decision making agents, and do not have believes nor desires. Decision making agents need to work with other agents in EOC through collaboration to deliver resources information and to deploy actions. Decision making agents need to make decisions to control the spread of pandemic influenza using available effective resources. Decision making agents need to determine the course of actions to be deployed to control the pandemic.

**Table 5.** Agent roles and number of agents per EOC

| Agent Role | Responsibilities | # of Agents |
|---|---|---|
| Decision Making Agent | Providing course of actions | 1 |
| | Adding new cases to memory | |
| Tactical Communication Agent | Receiving reports from operational agents | 1 |
| | Receiving resources reports from tactical agents | |
| | Sending reports to decision making agent | |
| | Receiving plan from decision making agent | |
| | Assigning plan tasks to tactical agents | |
| | Receiving status of plan execution | |
| | Sending plans status to decision making agent | |
| Tactical Agent | Handling reports from communication agent | 1 or more |
| | Processing available resources | |
| Operational Agent | Reporting current environment state | 1 or more |
| | Deployment of course of actions | |

**Design of the AIS Planning Methodology in the Proposed Response Model**
Design of AIS planning methodology follows Stepney et al. [28] structure of AIS engineering. Pandemic situation is represented as a record of the total number of agents based on agents' health states. For example, situation can be represented by: (Susceptible: 50 agents, In-Contact: 10 agents, Infectious: 20 agents, Isolated Infected: 10 agents, Recovered: 1 agent, and Dead: 3 agents). The required course of actions to control given pandemic situation is represented as follows: (identifier, successfulness of the course of actions, and current pandemic situation). In addition the course of actions is coupled with deployed actions. The entry of course of actions and its coupled actions constructs a memory case.

The city block distance is used here to find the similarity between situations. For example: distance between situation 1 (Infectious: 2 agents, Isolated Infectious: 1 agent) and situation 2 (Immunized: 31 agents) is: $|0-2|+|0-1|+|31-0|=33$.

Immune algorithms present the mutation (dynamic) behavior of AIS. Bone marrow algorithm, positive selection algorithm, clonal selection is used in the proposed model implementation.

## 4  Experiments

The main goal of experiments scenarios is to validate the proposed response model. Scenarios basically include simulation of pandemic influenza in a closed population of 1000 agents and initially three infected agents located in Cairo. Cairo EOC contains 3 operational agents, 2 tactical agents, and one decision making agent. The duration of the simulation round is 50 days. All simulation rounds involve randomly generated resources pool. Each action in the resource pool has efficacy of (0.75). The basic simulation round with no control strategies gives pandemic peak on day 10 with %60.8 infected agents [19].

Fig. 4 shows the flow of control during a simulation round, and the total cost of deployed actions (total cost = 2821.4 and plan certainty = 0 due to there are no previous plans in the system). While, Fig. 5 shows the stored case memory for the finished simulation round (case id = 174, successfulness = 0.001198). It is found that the pandemic peak is shifted to day 16 with %55 infected agents of the population.

**Fig. 4.** Round 1 - Crisis Response Log

| Case ID | Successfulness | S | E | I | II | R | IM | D |
|---------|----------------|---|---|---|----|---|----|----|
| 174 | 0.01198 | 0 | 0 | 3 | 1 | 0 | 0 | 0 |

| Action Type | From Date | To Date |
|-------------|-----------|---------|
| TARGETED_SOCIAL_DISTANCING | 7 | 38 |
| MASS_VACCINATION | 38 | 50 |
| QUARANTINING | 29 | 31 |
| AWARENESS | 42 | 50 |
| AWARENESS | 6 | 33 |
| QUARANTINING | 26 | 39 |
| TARGETED_SOCIAL_DISTANCING | 26 | 29 |
| TARGETED_SOCIAL_DISTANCING | 37 | 38 |
| TARGETED_SOCIAL_DISTANCING | 8 | 22 |
| QUARANTINING | 4 | 37 |
| TARGETED_VACCINATION | 28 | 37 |
| MASS_SOCIAL_DISTANCING | 30 | 46 |
| MASS_SOCIAL_DISTANCING | 2 | 37 |

**Fig. 5.** Round 1 - Case Memory

## 5   Conclusions

Currently, multi-agent architecture is the essence of response systems. The original idea comes out from agent characteristics in MAS, such as autonomy, local view of environment, capability of learning, planning, coordination and decentralized decision making. The incorporation of multi-agent systems can be clarified by discussing major disciplines involved during crisis response operations; situation management, and decision making and planning.

The study of biological systems is of interest to scientists and engineers as they turn out to be a source of rich theories. They are useful in constructing novel computer algorithms to solve complex engineering problems. Immunology as a study of the immune system inspired the evolution of artificial immune system, which is an area of vast research over the last few years. Detecting non-self, matching receptors, clone antibodies, response then detecting non-self cells again represents the control loop in the AIS metaphor. This process loop (control loop) is the core of the decision making process in AIS metaphor. The AIS model allows learning by mutating genes to generate acceptable receptors and store them in memory cells. Selection of required antibodies to be cloned and the cloning process represents the planning methodology of AIS. Artificial immune systems represent an interesting metaphor for building effective based and defensive multi-agent crisis response operations model. The proposed architecture proposed by AIS, promises effective operations and system architecture for crisis response.

According to experiments scenarios, AIS model shows slow learning process but it is very fast in handling desired or un-desired situations. Number of cases represents the growth of the system. While, scenarios results show that the effectiveness of response operations and utilization of resources are improved within the growth of the response model by ignoring low successfulness memory cases while deliberating new course of actions. The simulation process is very slow and consumes a lot of computation power. Each round takes at least 10 minutes to complete. It is recommended to implement the model using high performance computing to enable fast growth of case memory.